# Reevaluating Specificity in Neuroimaging: Implications for the Salience Network and Methodological Rigor

Tommaso Costa[1,2,3] & Franco Cauda[1,2,3]

[1]GCS-fMRI, Koelliker Hospital and Department of Psychology, University of Turin, Turin, Italy.
[2]FOCUS Laboratory, Department of Psychology, University of Turin, Turin, Italy.
[3]Neuroscience Institute of Turin (NIT), Turin, Italy.

**Abstract**

The accurate assessment of neuroimaging specificity is critical for advancing our understanding of brain disorders. Current methodologies often rely on frequentist approaches and limited cross-pathology comparisons, leading to potential overestimations of specificity. This study critiques these limitations, highlighting the inherent shortcomings of frequentist methods in specificity calculations and the necessity of comprehensive control conditions. Through a review of the Bayesian framework, we demonstrate its superiority in evaluating specificity by incorporating probabilistic modeling and robust reverse inference. The work also emphasizes the pivotal role of well-defined control conditions in mitigating overlap among brain pathologies, particularly within shared networks like the salience network. By applying Bayesian tools such as BACON (Bayes fACtor mOdeliNg), we validate the ability to derive disease-specific patterns, contrasting these with the narrower findings of frequentist analyses. This paper underscores the importance of Bayesian methodologies and extensive meta-analytic datasets in overcoming existing challenges, ultimately paving the way for more precise neuroimaging studies.

**Introduction**
The investigation of abnormalities in white and gray matter associated with various brain disorders has been a central focus of neuroimaging research. Over the past decade, numerous studies have aimed to identify specific neuroanatomical patterns linked to conditions such as schizophrenia, substance use disorders, and neurodegenerative diseases. However, a persistent challenge in this field is the substantial heterogeneity of findings across studies, often attributable to methodological limitations, including the reliance on frequentist approaches and incomplete cross-pathology comparisons.

For instance, a recent article published in this journal, *"Heterogeneous patterns of brain atrophy in schizophrenia localize to a common brain network"*, explored neuroimaging abnormalities in schizophrenia and claimed to assess their specificity by contrasting atrophy patterns with a limited selection of other brain disorders (Makhlouf et al., 2024). Similarly, Stubbs et al. (2023) investigated substance use disorders and argued for the specificity of a shared brain network by comparing it with atrophy patterns observed in aging and neurodegenerative conditions. Both studies adopted the methodology previously utilized by Darby et al. (2019) to derive specificity in neuroimaging findings.

While these studies represent significant progress in understanding network-level localization, their approach to assessing specificity has at least two notable limitations. Specifically: i) Poorly Defined Control Conditions for Measuring Specificity: they rely on comparisons with a narrowly defined subset of conditions, rather than accounting for the full range of disorders that may produce overlapping damage in the same brain regions. This constrained methodology raises important concerns, as it risks overestimating the uniqueness of the observed findings. ii) Limitations of the Frequentist Approach in Measuring Specificity: They utilized a frequentist framework which, as we will discuss later, is not well-suited to identifying true specificity. It is important to highlight that Tal Yarkoni et al. (2011), in their initial validation of Neurosynth, similarly claimed to compute reverse inference, equating it with specificity. However, Yarkoni and colleagues (Yarkoni, 2015) later revised this claim, acknowledging that the frequentist methodology employed did not provide a true measure of specificity. Consequently, they updated their framework to more accurately reflect its limitations.

**The Importance of Bayesian Approaches**
The significance of Bayesian approaches in calculating specificity has been thoroughly demonstrated by Cauda and Costa. In their seminal work, *"Finding Specificity in Structural Brain Alterations Through Bayesian Reverse Inference"*, Cauda et al. (2020) validated a fully Bayesian method for calculating specificity. This approach demonstrated its ability to generate disease-specific maps for conditions such as schizophrenia and Alzheimer's disease. To make this methodology widely accessible, they later introduced BACON (*Bayes fACtor mOdeliNg*), a freely available tool that allows researchers to calculate specificity directly from meta-analytical data. This tool enables robust reverse inference in both structural and functional neuroimaging studies (Costa et al., 2021). In contrast, the article by Makhlouf et al. (2024), while conducting a specificity analysis, employed a frequentist approach that is not well-suited for assessing specificity (see also the debate regarding the frequentist approach to reverse inference in Neurosynth). Indeed the statistical analyses performed by Makhlouf and colleagues rely primarily on classic t-tests, supplemented by a randomization process to assess statistical significance (Makhlouf et al., 2024). In addition, to generate specific maps, they used t-values exceeding the significance threshold. By applying different thresholds, they argued that their results were robust and reliable.

Frequentist statistics, which combines ideas from Fisher and Neyman-Pearson frameworks (Perezgonzale, 2015), is rooted in the concepts of the p-value and decision thresholds. The p-value is defined as the probability of observing an outcome as extreme as, or more extreme than, the one obtained in the data, assuming the null hypothesis ($H_0$) is true. Mathematically, it is expressed as:

$$p\text{-value} = P(t_{exp} > t_{crit} \mid H_0 \text{ is true}).$$

Within this framework, the analysis focuses exclusively on the null hypothesis ($H_0$), without explicitly considering the alternative hypothesis ($H_1$). Consequently, the p-value does not convey information about the specificity of a phenomenon, understood as its unique or exclusive characteristics. Moreover, since the decision relies on a critical threshold, all values exceeding this threshold are treated equivalently. This means that it is not possible to assert whether a t-value above the critical threshold is "better" or "worse" than another value that also surpasses the threshold.

In addition, inferential analyses in frequentist statistics are limited to evaluating the null hypothesis and provide no insight into alternative hypotheses. Even non-significant results cannot be confidently interpreted as evidence of no effect. In other words, the failure to reject the null hypothesis does not guarantee the absence of an effect (Wagenmakers, 2007; Dienes, 2014).

To address these limitations, evaluating specificity requires comparing two hypotheses: one representing specificity and the other representing non-specificity. This can be achieved effectively within the Bayesian framework using the Bayes Factor (BF), which quantifies the relative evidence for two competing hypotheses:

$$BF_{01} = \frac{P(H_0|D)}{P(H_1|D)}.$$

where $D$ represents the observed data.

The distinction from frequentist statistics lies in this comparative approach. In the Bayesian framework, it is necessary to account for all alternative scenarios unrelated to the phenomenon under study to make a probabilistic judgment. Frequentist methods, on the other hand, provide information solely about the probability of observing more extreme values than those obtained, based on hypothetical outcomes.

From the Bayes Factor, we can derive the probability of specificity using a straightforward transformation:

$$BF(H_0|D) = \frac{BF_{01}}{BF_{01} + 1}$$

This approach enables the computation of posterior probabilities directly from the Bayes Factor. Consequently, researchers can select a probability threshold to decide whether to accept or reject a hypothesis, offering greater flexibility and interpretability compared to frequentist methods.

**The Importance of Well-Defined Control Conditions in Assessing Specificity**
The paper "The Alteration Landscape of the Cerebral Cortex" by Cauda et al. (2019) revealed that structural brain alterations caused by diverse pathologies frequently converge in the same regions, significantly reducing the specificity of forward inference methods. Using entropy metrics, the study demonstrated that most brain pathologies exhibit substantial overlap within specific brain networks, such as the salience network. Conversely, by applying negentropy metrics to voxel-based morphometry (VBM) data from large-scale meta-analyses, the authors identified brain regions with low alteration diversity, which are more likely to reflect disease-specific changes.

The observation that most brain pathologies result in widespread brain alterations converging within the same regions necessitates the use of specific techniques to accurately calculate specificity. This calculation requires statistical comparisons between the brain alterations caused by a specific pathology and those produced by all other brain pathologies that may potentially affect the same regions. The findings by Cauda et al. in their paper "The Alteration Landscape of the Cerebral Cortex" demonstrate that nearly all brain pathologies cause at least some brain alterations within a set of regions broadly associated with the salience network. As a result, accurately computing specificity requires comparing two hypotheses: one representing specificity and the other representing non-specificity. This necessitates *considering all pathologies that share at least one common area of brain alteration*. Consequently, a comprehensive evaluation of brain specificity demands the inclusion of all known brain pathologies. Since this is practically unfeasible, the minimum requirement is to incorporate all studies investigating brain pathologies currently accessible in major meta-analytic databases.

The approach used in previous studies—comparing a given pathology with a small subset of other brain pathologies that may affect the same regions—is inherently flawed and leads to inaccurate results. This limitation is evident when comparing the findings of the paper "Heterogeneous Patterns of Brain Atrophy in Schizophrenia Localize to a Common Brain Network" by Makhlouf et al. (2024) with those of Cauda et al. in "Finding Specificity in Structural Brain Alterations Through Bayesian Reverse Inference" (2020). In their Bayesian approach, Cauda et al. included all pathologies available in the BrainMap database at the time. Their findings, in contrast to those of Makhlouf et al., revealed no specific alterations within the salience network. This result aligns with the evidence presented in "The Alteration Landscape of the Cerebral Cortex", which showed that the majority of overlapping brain alterations caused by various pathologies are located broadly within the salience network.

These findings underscore the critical importance of selecting well-defined and comprehensive control conditions when assessing specificity. Without such thorough consideration, specificity results risk being severely compromised.

**Conclusion:**

Neuroimaging has made substantial progress, but two critical issues remain in assessing specificity: the limitations of frequentist methods and the need for comprehensive control conditions. Frequentist approaches are less suited to calculating specificity compared to Bayesian methods, which leverage probabilistic modeling to account for shared and unique brain alterations. Additionally, accurate specificity analysis requires comparing the target pathology against all relevant conditions to avoid misleading conclusions.

Bayesian frameworks, such as those applied by Cauda et al. (2020), reveal true disease-specific alterations while addressing these challenges. By adopting Bayesian methodologies and ensuring well-defined control conditions, future research can achieve more accurate localization and a deeper understanding of brain disorders.

# Citations